\documentclass[letterpaper]{article} 
\usepackage{aaai25}  
\usepackage{times}  
\usepackage{helvet}  
\usepackage{courier}  
\usepackage[hyphens]{url}  
\usepackage{graphicx} 
\usepackage{subfigure}
\usepackage{natbib} 
\usepackage{caption}
\usepackage{algorithm}
\usepackage{algorithmic}

\usepackage{amsmath} 
\usepackage{amsfonts} 
\usepackage{booktabs}
\usepackage{multirow}
\usepackage{caption}
\usepackage{graphicx}
\usepackage{newfloat}
\usepackage{listings}
\usepackage[misc]{ifsym}

\DeclareCaptionStyle{ruled}{labelfont=normalfont,labelsep=colon,strut=off} 
\floatstyle{ruled}
\newfloat{listing}{tb}{lst}{}
\floatname{listing}{Listing}

\nocopyright

\title{Improving EEG Classification Through Randomly Reassembling Original and Generated Data with Transformer-based Diffusion Models}

\author{
    Mingzhi Chen,
    Yiyu Gui,
    Yuqi Su,
    Yuesheng Zhu,
    Guibo Luo$^{\ast}$,
    Yuchao Yang$^{\ast}$
}
\affiliations{
    \textsuperscript{}School of Electronic and Computer Engineerng, Peking University\\

    luogb@pku.edu.cn, yuchaoyang@pku.edu.cn
}

\usepackage{bibentry}

\begin{document}

\maketitle

\begin{abstract}
Electroencephalogram (EEG) classification has been widely used in various medical and engineering applications, where it is important for understanding brain function, diagnosing diseases, and assessing mental health conditions. However, the scarcity of EEG data severely restricts the performance of EEG classification networks, and generative model-based data augmentation methods have emerged as potential solutions to overcome this challenge. There are two problems with existing methods: (1) The quality of the generated EEG signals is not high; (2) The enhancement of EEG classification networks is not effective. In this paper, we propose a Transformer-based denoising diffusion probabilistic model and a generated data-based augmentation method to address the above two problems. For the characteristics of EEG signals, we propose a constant-factor scaling method to preprocess the signals, which reduces the loss of information. We incorporated Multi-Scale Convolution and Dynamic Fourier Spectrum Information modules into the model, improving the stability of the training process and the quality of the generated data. The proposed augmentation method randomly reassemble the generated data with original data in the time-domain to obtain vicinal data, which improves the model performance by minimizing the empirical risk and the vicinal risk. We verify the proposed augmentation method on four EEG datasets for four tasks and observe significant accuracy performance improvements: 14.00\% on the Bonn dataset; 6.38\% on the SleepEDF-20 dataset; 9.42\% on the FACED dataset; 2.5\% on the Shu dataset. We will make the code of our method publicly accessible soon.
\end{abstract}

\section{Introduction}

Electroencephalography (EEG) records the electrical wave changes in brain activity, providing a comprehensive reflection of the electrophysiological activity of brain neurons. Currently, EEG classification has been widely applied in various medical and engineering applications, such as seizure detection, emotion recognition, sleep stage classification, motor imagery, etc \citep{1,2,3,4}. However, the limited amount of EEG data severely restricts the performance of EEG classification networks, and generative model-based data augmentation methods have shown promise as effective solutions.

\citep{5,6} used Generative Adversarial Network (GAN) to generate EEG data to improve the performance of EEG classification networks. But the GAN-based networks are prone to mode collapse during training, making it difficult to guarantee the quality of the generated EEG signals. \citep{7,8} utilized the diffusion model to generate EEG data in order to increase the size of the training dataset, and they respectively validated their methods on seizure detection and emotion recognition tasks. However, \citep{7,8} only validated their methods on a single task, without assessing the applicability of the proposed methods on a wider range of EEG tasks. Additionally, the model structures they proposed are complex and lack scalability. 

Previous data augmentation methods for EEG classification networks based on generative models often directly incorporate generated data into the training dataset. However this method did not significantly improve the performance of EEG classification networks and could even lead to performance degradation. The primary reason for this is the lack of reconstruction of the labels for the generated EEG data, which can easily lead to the model learning incorrect information under the condition of minimizing empirical risk \citep{9}, thereby affecting the model's performance.

Overall, the above methods have three main issues: (1) The quality of the generated EEG signals by the previous models is difficult to effectively guarantee. (2) Directly incorporating generated data into the training data may not effectively improve the performance of EEG classification networks. (3) The previous methods have not been validated across multiple different tasks to confirm their effectiveness.

\begin{figure*}[!ht]
    \centering
    \includegraphics[width=0.95\textwidth]{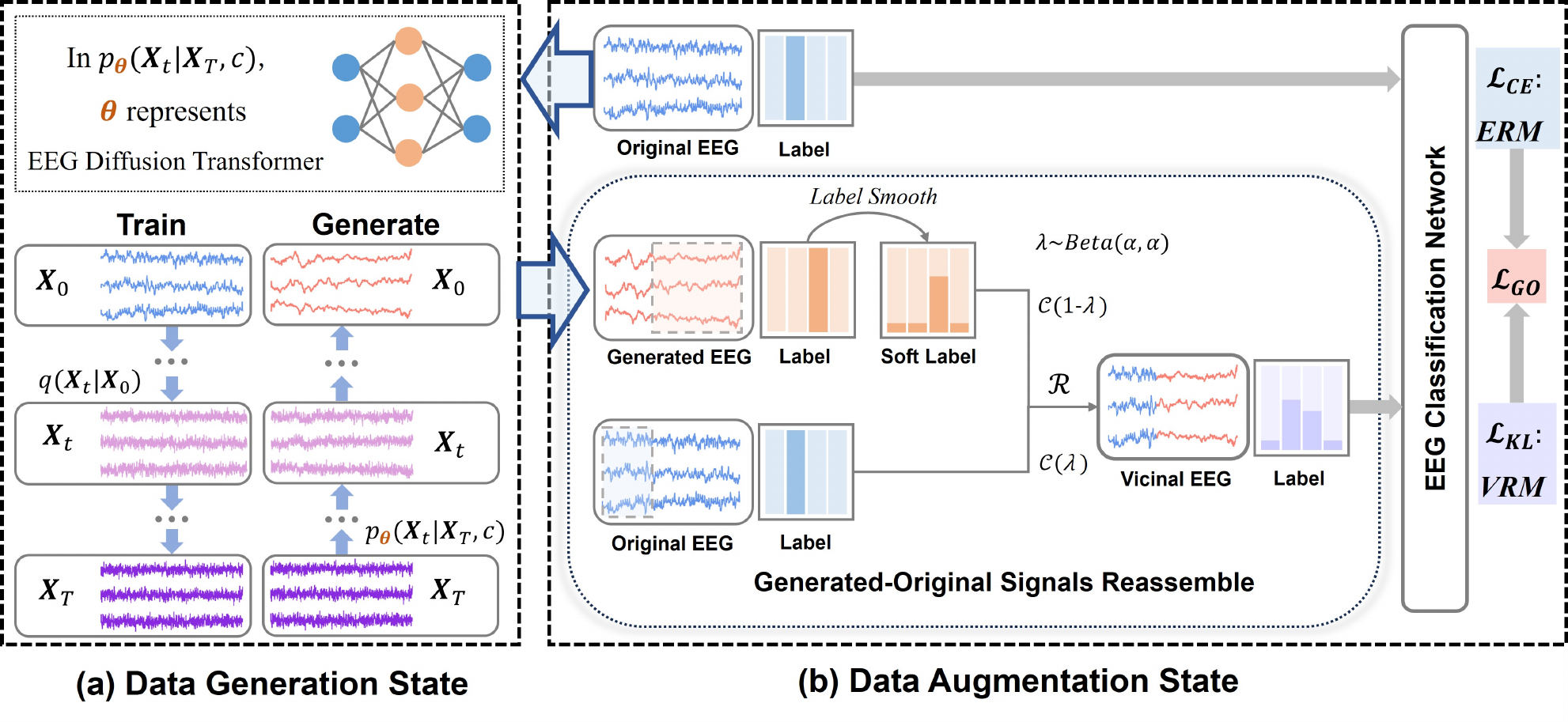}
    \caption{The Illustration of the Proposed Method.}
    \label{Figure1}
\end{figure*}

In order to improve the quality of generated EEG signals, we improved the traditional diffusion model network structure for the characteristics of EEG signals. Compared to the diffusion model structures in \citep{7,8}, our model features a simpler architecture that is more extensible and capable of generating high-quality EEG signals. To effectively improve the performance of EEG classification networks, we propose an effective data augmentation method based on generated data. The proposed augmentation method randomly reassembles the generated data with the original data to obtain the vicinal data, which improves the model performance by minimizing the empirical risk and the vicinal risk \citep{10}. We validated our proposed method on EEG datasets from four different tasks: seizure detection, emotion recognition, sleep stage classification, and motor imagery. Notably, we observed significant performance improvements across all datasets, indicating the outstanding universality of our proposed method. 
Our contributions are as follows:

\begin{itemize}
    \item We propose a Transformer-based denoising diffusion probabilistic model, which incorporated Multi-Scale Convolution and Dynamic Fourier Spectrum Information modules, and is able to generate high-quality EEG signals.
    \item We propose a generated data-based data augmentation method, which randomly reassembles generated and original EEG data in the time-domain to obtain vicinal data and subsequently improves model performance by minizing the empirical risk and the vicinal risk.
    \item By evaluating our method on four widely used EEG datasets from diverse tasks, we demonstrate that it surpasses the baseline in both the quality of generated EEG signals and the effectiveness of data augmentation.
\end{itemize}

\section{Related Work}
\subsection{Generative Models} In recent years, there has been rapid development in generative models, including Variational Autoencoders (VAEs) \citep{35,36}, Flow models \citep{38}, Generative Adversarial Networks (GANs) \citep{39,40,41}, and Denoising Diffusion Probabilistic Models (DDPMs) \citep{11,12,42,43}. VAEs use an encoder to map input data to a latent Gaussian distribution and a decoder to reconstruct the data. Flow models transform data through reversible steps into a simple prior, like a Gaussian, and sample by transitioning from this prior to the target distribution. GANs have a generator creating fake data and a discriminator distinguishing real from fake, trained adversarially. DDPMs add noise to data progressively to reach a standard normal distribution and generate samples by gradually denoising.

\subsection{EEG Data Augmentation via Generative Models} 
\citep{5,6} have leveraged Generative Adversarial Networks (GANs) to generate EEG data, aiming to enhance the classification capabilities of EEG-based networks. Despite the potential of GANs, they are susceptible to mode collapse, a challenge that can degrade the quality of the generated EEG signals. The emergence of Denoising Diffusion Probabilistic Models\cite{11} has provided a more stable model for generating EEG signals compared to GANs. \citep{7,8} utilized diffusion models to generate EEG data in order to expand the scale of the training dataset, and they have validated their methods on tasks related to seizure detection and emotion recognition, respectively. However, the scope of their validation was limited to these specific tasks, without exploring the broader applicability of their methods across a variety of EEG applications. Moreover, the complexity of the models they introduced may hinder their scalability in diverse EEG-related tasks.

\section{EEG Data Generation and Augmentation}

Our proposed method consists of two stages: the first is the diffusion model-based data generation stage, and the second is the generated data-based data augmentation stage, as depicted in Figure \ref{Figure1}.

In the first state: We progressively add noise into the original EEG signals and utilize our proposed diffusion model to predict the the noise added. Finally, we utilized the well-trained diffusion model to generate EEG data through a multi-step denoising.

In the second state: The labels of the generated EEG data were reconstructed through the process of label smoothing. Subsequently, both the original and generated data were randomly segmented and reassembled to generate vicinal data. The original and vicinal data were inputted to the EEG classification network, where the network performance was enhanced through optimization of the cross-entropy loss for original data and the KL divergence loss for vicinal data.

\subsection{EEG Data Generation}
\subsubsection{Denoising Diffusion Probabilistic Models}
In this paper, we utilize the denoising diffusion probabilistic model (DDPM) \citep{11} to generate EEG data. DDPM learns the data distribution by modeling the diffusion process (i.e., adding noise process) and generates new data by gradually denoising through the reverse process.

Define $x_0\in\mathbb{R}^{C\times L}$ as the raw EEG data, $x_t\in\mathbb{R}^{C\times L}$ represents the noised EEG data at step \textit{t}, and $q(x_0)$ as the EEG data distribution, where \textit{C} is the EEG channel and \textit{L} is length of EEG segment. Its forward diffusion process is defined as a fixed Markov process:

\begin{equation}
    q(x_{1:T}|x_0)=\prod_{t=1}^{T}{q(x_t|x_{t-1})}
\label{eq1}
\end{equation}
\begin{equation}
    q(x_{t-1}|x_t):=N(\sqrt{1-\beta_t}x_{t-1},\beta_tI)
\label{eq2}
\end{equation}

Where $x_0$ represents the EEG data, and $\beta_t$ represents the noise level, and through reparameterization trick, we can directly sample from $x_t$:
\begin{equation}
    x_t = \sqrt{\bar{\alpha_t}}x_0+\sqrt{1-\bar{\alpha_t}}\epsilon
\label{eq3}
\end{equation}

Where $\alpha_t=1-\beta_t$ , $\bar{\alpha_t}=\prod_{t=1}^{T}\alpha_{t\ }$, $\ \epsilon \sim N(0,I)$. The reverse process is defined as:

\begin{equation}
    p_\theta(x_{0:T})=p(x_{T})\prod_{t=1}^{T}{p_\theta(x_{t-1}|x_t)}
\label{eq4}
\end{equation}
\begin{equation}
    p_\theta(x_{t-1}|x_t):=N(\mu_\theta(x_t,t),\sigma_tI)
\label{eq5}
\end{equation}

\begin{figure*}[!ht]
    \centering
    \includegraphics[width=0.85\textwidth]{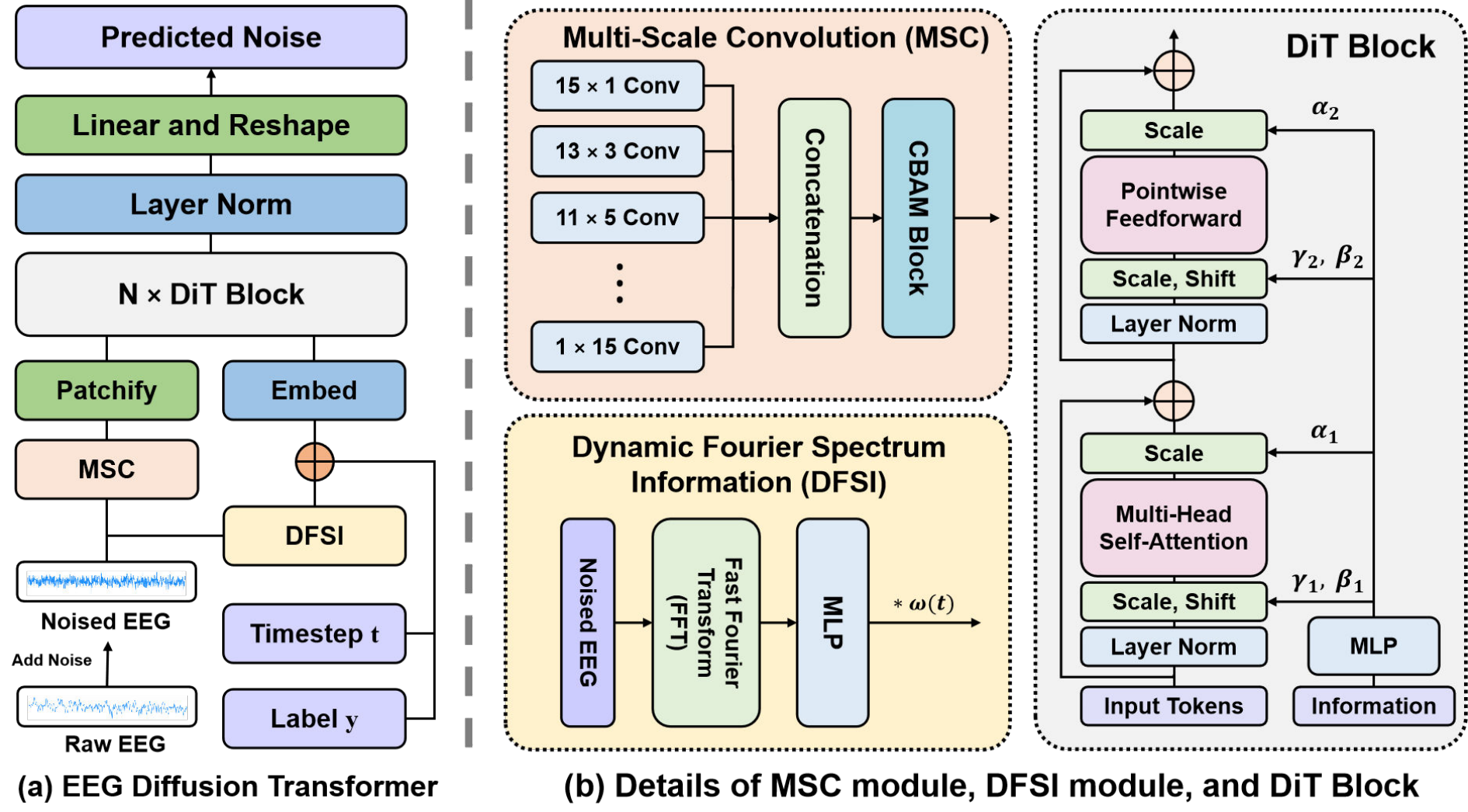}
    \caption{(a) depicts the overall architecture of EEG Diffusion Transformer, while (b) provides specific details of the Multi-Scale Convolution (MSC) module, Dynamic Fourier Spectrum Information (DFSI) module, and Diffusion Transformer (DiT) block.}
    \label{Figure2}
\end{figure*}

Where $p(x_T) \sim N(0,I)$ , $\sigma_t=\frac{1-\bar{\alpha_{t-1}}}{1-\bar{\alpha_t}}\beta_t$. To approximate the original reverse process $q(x_{t-1}|x_t)$, the model is trained by maximizing the approximate value of the evidence lower bound (ELBO) of the log-likelihood.By utilizing a neural network $\epsilon_\theta$ as a noise predictor to predict the added noise $\epsilon$, the overall training objective becomes: 
\begin{equation}
    E_{x,\epsilon \sim N(0,I),t}[||\epsilon_{\theta}(\sqrt{\bar{\alpha_t}}x_0+\sqrt{1-\bar{\alpha_t}}\epsilon,t)-\epsilon||^2]
\label{eq6}
\end{equation}
For training conditional diffusion models, conditioning information \textit{c} can be further integrated into the model, and the training objective becomes:
\begin{equation}
    E_{x,\epsilon \sim N(0,I),t}[||\epsilon_{\theta}(\sqrt{\bar{\alpha_t}}x_0+\sqrt{1-\bar{\alpha_t}}\epsilon,t,c)-\epsilon||^2]
\label{eq7}
\end{equation}

\subsubsection{EEG Data Preprocessing}
The diffusion model assumes that after \textit{T} iterations of diffusion, the noise-contaminated data $x_{T}$ approximately follows a standard normal distribution, i.e., $q(x_T)\approx N(0,I)$. Under the premise of satisfying this assumption, data can be generated through the reverse process, which requires limiting the range of the original data.

Whether it is the original image diffusion model \citep{11} or the implicit diffusion model \citep{34} for image compression using variational autoencoders (VAEs) \citep{35,36}, the value of the original data is limited within a certain range. Thus, it is easy to satisfy the assumption that $q(x_T)\approx N(0,I)$.

However, for EEG data, the amplitude of signals generally does not have numerical range constraints, which can easily lead to the non-convergence of the diffusion model during training. A common approach is to utilize z-score normalization to reduce the amplitude of EEG signals. However, this approach leads to the loss of mean and standard deviation information of the processed EEG data, and this loss is irreversible.

Therefore, this paper proposes a simple and effective preprocessing method for signal data that preserves as much information as possible under the assumption of the diffusion model. The method selects a constant-factor and then divides the signal amplitude by this constant-factor, aiming to scale the amplitude to the range [-4, 4]. For the generated data from the diffusion model, it is necessary to multiply the generated data by the constant-factor used in training to restore the signal amplitude information.

\subsubsection{EEG Diffusion Transformer}

We use the proposed EEG Diffusion Transformer model as a noise predictor network $\epsilon_\theta$ for EEG data generation, and the overall architecture of the proposed model is depicted in Figure \ref{Figure2}. The EEG Diffusion Transformer is primarily composed of a Multi-Scale Convolution (MSC) module, a Dynamic Fourier Spectrum Information (DFSI) module, and multiple Diffusion Transformer \citep{12} (DiT) blocks. 

The noised EEG is first processed by the MSC module to extract both low-frequency and high-frequency features. Meanwhile, the DFSI module extracts the Fourier spectrum information of the noised EEG data. Subsequently, a convolutional layer is utilized to segment the features into patches, resulting in a sequence of tokens. The guiding information is constituted by the sum of the Fourier spectral information, the categorical information, and the temporal information. The token sequence, along with the guiding information, is processed through multiple DiT blocks. The model ultimately outputs the predicted noise.

\textbf{Multi-Scale Convolution Module} EEG signals can be broadly categorized into five frequency bands \citep{13}, \citep{14,15,16} indicating that distinct physiological activities are characterized by different EEG frequency band signatures. The frequency diversity in EEG signals challenges single-sized convolutional kernels. Thus, we use the Multi-Scale Convolution (MSC) module for effective feature extraction. This module, comprising convolutional kernels of varying sizes, enables the extraction of both low-frequency and high-frequency features from the EEG signals. Subsequently, the low-frequency and high-frequency features are subjected to adaptive feature refinement via a Convolutional Block Attention Module (CBAM) \citep{17}.

\textbf{Dynamic Fourier Spectrum Information Module} For conditional EEG diffusion models, utilizing class information as a guiding information is a common way to guide the model in generating EEG signals. However, for unconditional diffusion models, the absence of class guidance slow down the convergence rate, thereby affecting the generation performance of the model. The Fourier spectrum can reflect numerous information about the EEG in the frequency domain, and it can be used as guiding information to guide the model in generating EEG signals. Therefore, to enhance the generation performance of unconditional EEG diffusion models, we propose a Dynamic Fourier Spectral Information (DFSI) module. Specifically, the noised EEG signal is first subjected to a Fast Fourier Transform, followed by the application of a Multi-Layer Perceptron (MLP) to encode the features of the Fourier spectrum, thereby obtaining the Fourier spectral information, as shown in Equation \ref{eq8}, where $I_{FS}$ represents the Fourier spectrum information, \textit{FFT } represents the operation of Fast Fourier Transform on noised EEG signals $x_{t}$.
\begin{equation}
    I_{FS}=MLP(FFT(x_t))
\label{eq8}
\end{equation}
As the noise intensity increases, the information content contained in the Fourier spectrum of the noised EEG decreases. Therefore, we propose a time-step-dependent weight $\omega(t)$ to adjust the weight of the Fourier spectral information within the guiding information, which decreases as the time step increases.
\begin{equation}
    \omega(t)=cos(\frac{\pi}{2}\ast\frac{t}{T_{max}})
\label{eq9}
\end{equation}
For the unconditional diffusion model, the guiding information $I_{Un}$ consists of temporal information $I_{T}$ and Fourier spectral information $I_{FS}$, as shown in Equation \ref{eq10}. While for the conditional diffusion model, the guiding information \ $I_{Con}$ consists of class information $I_C$, temporal information $I_{T}$ and Fourier spectral information $I_{FS}$, as shown in Equation \ref{eq11}.
\begin{equation}
    I_{Un} = I_{T} + \omega(t)*I_{FS}
\label{eq10}
\end{equation}
\begin{equation}
    I_{Con} = I_{T} + I_{C} + \omega(t)*I_{FS}
\label{eq11}
\end{equation}

\subsection{Generated-Original Signals Reassembled Data Augmentation}

\subsubsection{Directly Incorporating Generated Data May Not Be a Good Idea}\citep{46} observed through extensive experiments that generated data is not always beneficial for representation learning and can sometimes be detrimental. For instance, adding DDPM-generated data directly to CIFAR-10 training led to a drop in classification accuracy by over 1\%. Similarly, \citep{47} finds that iterative training of the model with generated data leads to model collapse. The limited number of training datasets is insufficient to cover the entire data distribution, resulting in discrepancies between the data distribution generated by the model trained on the training set and the true data distribution. These discrepancies may lead the model to learn incorrect information, which in turn can cause a decline in model performance. 

Consider a classification model $\psi_{\theta}$  with trainable parameters $\theta$. Let $\mathbb{S} = \{(x^1,y^1),(x^2,y^2),...,(x^n,y^n)\}$ denote the training set of ${n}$ samples. The model is required to map the samples ${x}$ to labels $y$, with the training objective being to predict the posterior probabilities of the classes. The standard approach is to minimize the expected cross-entropy loss function $\mathcal{L}_{CE}$ over the training set:

\begin{equation}
    {\theta}^{*}=\mathop{\arg\min}\limits_{\theta}[\mathbb{E}_{x\sim\mathbb{S}}[\mathcal{L}_{CE}(\psi_{\theta},y)]]
\label{eq12}
\end{equation}

When the training set consists of real data set $\mathbb{S}_R = \{(x_R^1,y_R^1),(x_R^2,y_R^2),...,(x_R^P,y_R^P)\}$ and generated data set $\mathbb{S}_G = \{(x_G^1,y_G^1),(x_G^2,y_G^2),...,(x_G^P,y_G^Q)\}$, the model parameter updates can be obtained by solving the Equation \ref{eq13} using stochastic gradient descent with a learning rate $\lambda$ and a batch size $B$:

\begin{equation}
    \theta_{t+1}=\theta_{t}-\lambda\nabla_{\theta}\frac{1}{B}(\sum_{i=0}^p\mathcal{L}_{CE}(\psi_{\theta},y_{R}^i)+\sum_{j=0}^q\mathcal{L}_{CE}(\psi_{\theta},y_{G}^j))
\label{eq13}    
\end{equation}

where $p$ and $q$ denote the number of real samples and synthetic samples, respectively, in the same batch. When the batch size is sufficiently large, $\frac{p}{q} \approx \frac{P}{Q}$, indicating that the proportion of the gradient contribution to the model parameter updates is related to the ratio of real data to synthetic data in the training set. When the distribution of synthetic data deviates from the real data distribution, a higher proportion of synthetic data can lead the model to converge in a suboptimal direction.

\subsubsection{Label Reconstruction for Generated Data}
To mitigate the impact of erroneous information from generated data during the early stages of training, we reconstruct the labels of the generated data. Specifically, we apply Label Smoothing \citep{25} to increase the entropy of the generated data labels, as shown in Equation \ref{eq13}. $\beta$ represents a hyperparameter indicating the proximity between the generated data distribution and the original training data distribution. The closer the generated data distribution is to the original data distribution, the closer $\beta$ is to 1; otherwise, it is \textit{c}loser to 0. \textit{k} is the number of classification categories.

This method encourages the model to produce predictions with higher entropy when dealing with generated data, rather than low-entropy predictions as it would with real data, thus alleviating the misleading effects of generated data on the model.

\begin{equation}
{\widetilde{y}}_{gen}=OneHot(y_{gen})\ast\beta+\frac{(1-\beta)}{k}\ast\mathbf{1}
\label{eq14}
\end{equation}

\subsubsection{Obtain Vicinal Data by Reassembling Generated-Original Signals} The training of an EEG classification network is a process that leverages a neural network $f_{\theta}$ to learn the conditional distribution $P(Y|X)$ of EEG data X and their corresponding labels $Y$. The goal of the network is to minimize the expected loss function $\mathcal{L}$ on the training data, known as empirical risk minimization \citep{9}. However, when the training data size is small, minimizing empirical risk can easily lead to overfitting. Data augmentation based on vicinal data can alleviate this issue by minimizing vicinal risk\citep{10}. Vicinal data can be obtained through original data, such as directly adding noise to the original samples \citep{19}; mixing two sample data \citep{20,21,22,23}. 

Inspired by \citep{20,21}, We randomly reassemble original data and generated data to further enhance the diversity of training data distribution, as shown in Equation \ref{eq15}. Original data $x_{orig}$ and generated data $x_{gen}$ are randomly cropped and reassembled in the time dimension to produce a vicinal data $x_{vic}$. The label value of the vicinal data $y_{vic}$ is the sum of the labels of the two according to the cropping length ratio, as shown in Equation \ref{eq16}. $\lambda\sim Beta(\alpha,\alpha)$, $\lambda$ represents the proportion of original data cropped, $(1-\lambda)$ represents the proportion of generated data cropped, $\mathcal{C}$ represents the operation of signal cropping according to the proportion, and $\mathcal{R}$ represents the reassemble operation of signal segments.

\begin{equation}
    x_{vic}=\mathcal{R}(\mathcal{C}(\lambda, x_{orig}),\ \ \mathcal{C}((1-\lambda), x_{gen}))
\label{eq15}
\end{equation}
\begin{equation}
    y_{vic}=\lambda y_{orig}+(1-\lambda){\widetilde{y}}_{gen}
\label{eq16}
\end{equation}

\citep{22} proposes that merely minimizing vicinal risk can lead to models favoring high-entropy predictions. To address this problem, we explicitly combine vicinal risk with empirical risk while minimizing the combined risk. The overall loss function of EEG classification network $\mathcal{L}_{GO}$ termed Generated-Original Signal Reassembling (GO) loss function is expressed as the sum of the cross-entropy loss function $\mathcal{L}_{CE}$ for original data and the KL divergence loss function $\mathcal{L}_{KL}$ for vicinal data, as shown in Equation \ref{eq17}. In our experiments, both $\eta$ and $\alpha$ are set to 1.

\begin{equation}
    \mathcal{L}_{GO} = \mathcal{L}_{CE} (f(x_{orig}),y_{orig})+\eta\ast\mathcal{L}_{KL}(f(x_{vic}),y_{vic})
\label{eq17}
\end{equation}

The proposed GO loss function $\mathcal{L}_{GO}$ allows the EEG classification network to learn from a more diverse data distribution while being supervised by the distribution of original data.

\section{Experiments}

\subsection{EEG Classification Network}
The proposed data augmentation method is not limited to a specific network structure but is applicable to any EEG classification network. We select two different networks, EEGNet \citep{25} and EEG-Conformer \citep{26}, from classical EEG classification networks to validate the enhancement effect of our method. EEGNet is a classic EEG classification network based on depthwise separable convolutions, which has demonstrated stable and robust performance across a variety of EEG classification tasks. EEG-Conformer is an EEG classification network using spatiotemporal convolutions and Transformers to capture local and global features, performing well on public datasets for motor imagery and emotion recognition.

\subsection{Methods for evaluating the quality of generated EEG data}
The Fréchet Inception Distance (FID) is used to evaluate the quality of generated images in image generation models \citep{32}. FID is obtained by comparing the distributions of original and generated data in the embedding layer of a pre-trained network, typically InceptionV3. It assumes these distributions follow a multivariate Gaussian and calculates the Wasserstein-2 distance. A lower FID indicates higher image quality.

However, the calculation of FID for generated EEG signals lacks a pre-trained network similar to InceptionV3. EEG GAN\citep{5} approaches train an EEG classification network on the relevant EEG dataset and use this classification network to compute the FID score for the generated data. This paper adopts a similar approach to EEG GAN by using a classical EEG classification network, named EEG DeepConvNet \citep{37}, for FID computation. Considering variations in channel numbers, signal amplitudes, and sampling frequencies across EEG datasets, we trained four EEG DeepConvNets on each dataset to compute the FID for generated data. The features extracted from the final Max Pooling layer of these networks were then utilized to calculate the FID. Detailed training parameters are provided in the Experimental Setup section.

\subsection{Experimental Setup}
\subsubsection{Datasets} To compare the quality of EEG signals generated by different models and test the enhancement effects of various data augmentation methods on EEG classification networks, our experiments were conducted across four publicly available EEG datasets corresponding to four tasks: epilepsy detection, emotion recognition, sleep stage classification, and motor imagery. A detailed introduction to the datasets is provided in the Appendix, which includes four datasets: Bonn \citep{27}, SleepEDF-20 \citep{29}, FACED \citep{30}, and Shu \citep{31}. 

For the quality assessment of the generated signals, we trained the generative models on the entirety of the datasets and subsequently generated an equal quantity of EEG data, calculating the FID between the original and generated EEG data. It is noteworthy that since the evaluation of the quality of generated EEG data does not involve EEG classification networks, there is no risk of data leakage.

Regarding the enhancement effects on the classification network, the datasets were partitioned into training, validation, and testing sets with a ratio of 0.6:0.2:0.2. Generative models were trained solely on the training set, and these models were used to generate EEG data. Four random seeds were used to ensure the robustness of our results.

\subsubsection{Settings} For all EEG generative models and EEG classification networks, we uniformly employed a learning rate of 2e-4, with AdamW serving as the optimizer and a weight decay of 1e-6. The default number of training epochs for the generative models was set to 5,000, while for the classification networks, it was 1,000.

For all datasets, a unified batch size of 64 was utilized. To optimize both training efficiency and GPU memory usage, we selected three channels from the multi-channel datasets FACED and Shu. Our experiments revealed that employing these three channels can still achieve favorable classification outcomes and significantly expedite the training process of the generative model. All our experiments were conducted on a machine with Intel(R) Xeon(R) Platinum 8163 CPU @ 2.50GHz and a single NVIDIA GeForce 4090 GPU.

\subsection{Results}
\subsubsection{EEG Generation Quality}
We compared the FID of the EEG-GAN \citep{5}, TimeVAE \citep{33}, UNet-Diffusion \citep{8} and our proposed EEG Diffusion Transformer model on the generated data, as shown in Table \ref{Table1}. A lower FID indicates higher quality of the generated EEG signals.

\begin{table}[H]
\caption{Comparison of FID for EEG Signals Generated by Different Models}
\centering
\scalebox{0.9}{
\begin{tabular}{@{}ccccc@{}}

\toprule
Dataset            & Ours          & GAN   & VAE   & UNet-Diffusion \\ \midrule
Bonn               & \textbf{3.81} & 7.86  & 30.81 & 19.02          \\
SleepEDF-20        & \textbf{5.25} & 60.29 & 46.54 & 29.04          \\
FACED              & \textbf{0.55} & 18.24 & 11.33 & 2.01           \\
Shu                & \textbf{0.24} & 1.68  & 18.43 & 0.45           \\ \bottomrule
\end{tabular}
}
\label{Table1}
\end{table}

\begin{table*}[!ht]
\centering
\caption{Performance Comparison of Data Augmentation Methods on EEGNet and EEG-Conformer Across Various Datasets}
\scalebox{0.64}{
\begin{tabular}{lcccccccc}
\toprule
\multirow{2}{*}{Algorithm} & \multicolumn{4}{c}{EEGNet} & \multicolumn{4}{c}{EEG-Conformer} \\
\cmidrule(lr){2-5} \cmidrule(lr){6-9}
& Bonn & SleepEDF-20 & FACED & Shu & Bonn & SleepEDF-20 & FACED & Shu \\
\midrule
Original Performance & \textit{0.8125} $\pm$ \tiny{0.0830} & 0.7825 $\pm$ \tiny{0.0070} & \textit{0.6438} $\pm$ \tiny{0.0282} & \textit{0.5843} $\pm$ \tiny{0.0086} & \textit{0.8675} $\pm$ \tiny{0.0320} & 0.8238 $\pm$ \tiny{0.0044} & 0.6554 $\pm$ \tiny{0.0269} & \textit{0.6131} $\pm$ \tiny{0.0113} \\
Data Incorporation with TimeVAE & 0.7350 $\pm$ \tiny{0.0191} & 0.8005 $\pm$ \tiny{0.0104} & 0.6315 $\pm$ \tiny{0.0472} & \textit{0.5905} $\pm$ \tiny{0.0128} & 0.7925 $\pm$ \tiny{0.0411} & 0.8146 $\pm$ \tiny{0.0133} & 0.6837 $\pm$ \tiny{0.031} & 0.5981 $\pm$ \tiny{0.0074} \\
Data Incorporation with EEG GAN & 0.7725 $\pm$ \tiny{0.065} & 0.7780 $\pm$ \tiny{0.0081} & 0.6308 $\pm$ \tiny{0.0128} & 0.5491 $\pm$ \tiny{0.0274} & 0.7350 $\pm$ \tiny{0.0387} & \textit{0.8266} $\pm$ \tiny{0.0067} & 0.6533 $\pm$ \tiny{0.0398} & 0.6023 $\pm$ \tiny{0.0106} \\
Data Incorporation with EEG Diffusion Transformer & 0.8025 $\pm$ \tiny{0.0310} & \textit{0.8082} $\pm$ \tiny{0.0017} & 0.6428 $\pm$ \tiny{0.0210} & 0.5882 $\pm$ \tiny{0.0055} & 0.8500 $\pm$ \tiny{0.0408} & 0.8184 $\pm$ \tiny{0.0058} & \textit{0.7047} $\pm$ \tiny{0.0214} & 0.5921 $\pm$ \tiny{0.0113} \\
Our Data Augmentation Method & \textbf{0.9525} $\pm$ \tiny{0.0263} & \textbf{0.8463} $\pm$ \tiny{0.0040} & \textbf{0.7380} $\pm$ \tiny{0.0065} & \textbf{0.6093} $\pm$ \tiny{0.0098} & \textbf{0.9575} $\pm$ \tiny{0.0287} & \textbf{0.8625} $\pm$ \tiny{0.0023} & \textbf{0.7449} $\pm$ \tiny{0.0199} & \textbf{0.6370} $\pm$ \tiny{0.0018} \\
\bottomrule
\end{tabular}
}
\label{Table2}
\end{table*}
It is clear from Table \ref{Table1} that the FID of the EEG signals generated by our model are lower than those of the other three models on all four datasets. The results indicate that our model generates high quality EEG signals and achieves excellent generation results in different EEG tasks without relying on specific EEG datasets. On the Shu dataset, we observed that the quality of the EEG signals generated by our model is relatively close to that of the GAN and UNet-Diffusion models. However, on other datasets, such as the Bonn dataset, the FID of the signals generated by our model are significantly lower than those of the other models. 

\begin{figure}[H]
    \centering
    \subfigure[Spectrum of Shu Dataset]{
        \includegraphics[width=0.224\textwidth]{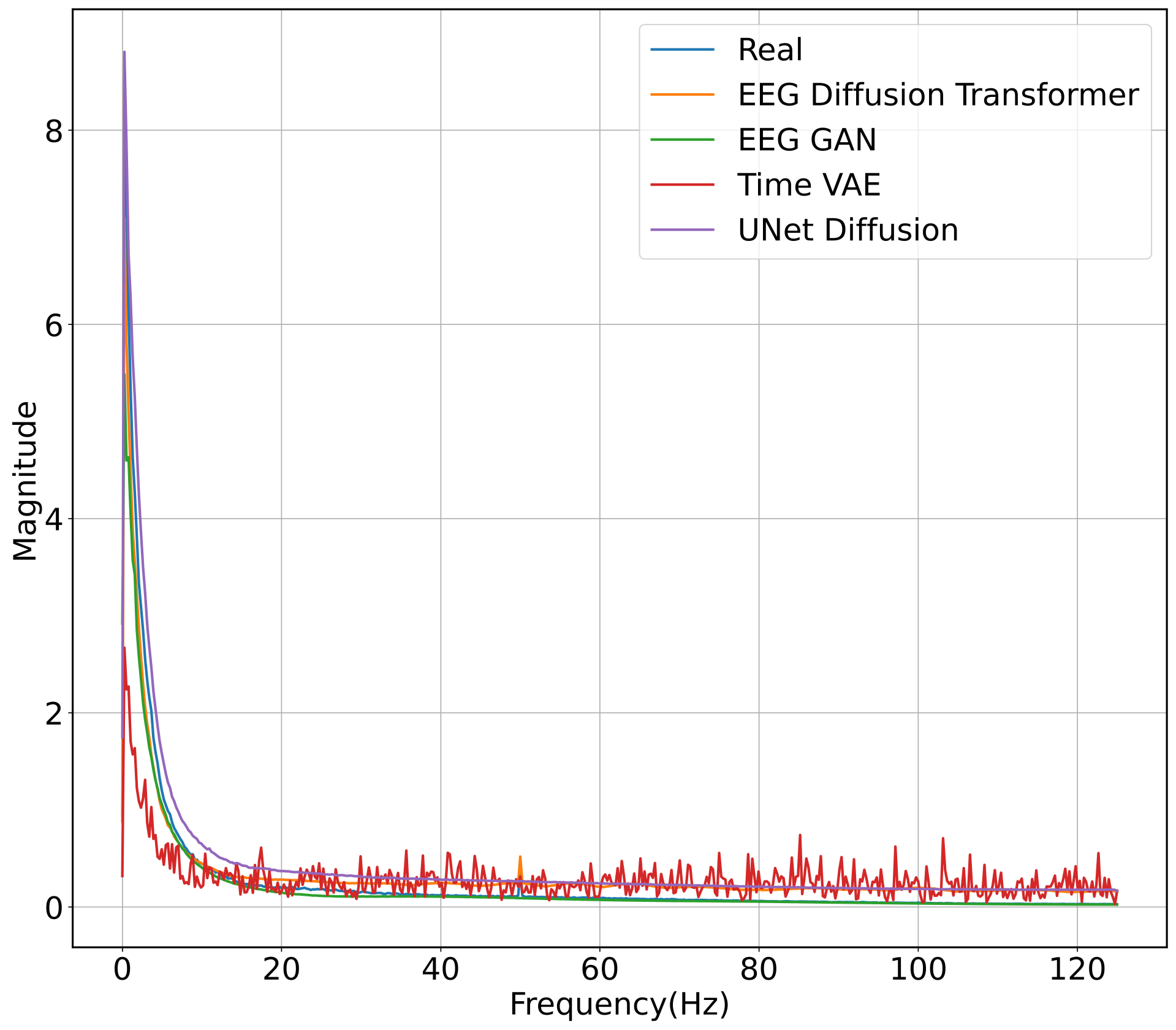}
    }
    \hfill
    \subfigure[Spectrum of Bonn Dataset]{
        \includegraphics[width=0.224\textwidth]{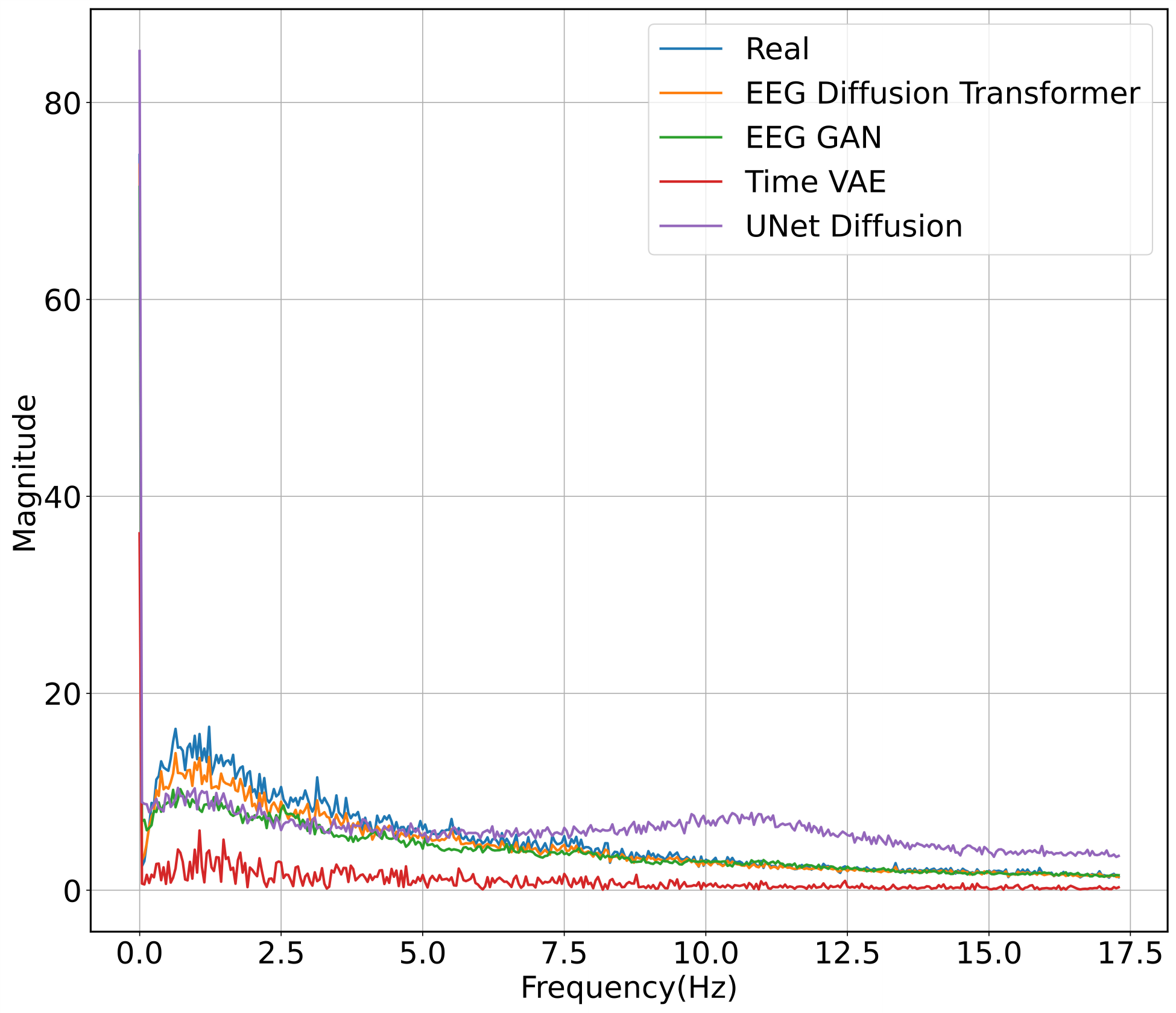}
    }
    \caption{Comparison of the Fourier Spectra of EEG Signals Generated by Different Models.}
    \label{Figure3}
\end{figure}

To explore the reasons for these differences, we compared the Fourier spectrum of the signals generated by different models with the original signals on the Shu dataset and the Bonn dataset, as shown in Figure \ref{Figure3}.

From Figure \ref{Figure3}, it is evident that the spectra of the Shu dataset are relatively simple, and most models generate signals whose spectra are quite close to the original. In contrast, the Bonn dataset has a more complex frequency composition. Except for our model and EEG GAN, the spectra generated by the other models significantly deviate from the original, resulting in poorer FID. The challenge of generating high-quality EEG signals is heightened by the complex frequency composition, yet our model demonstrates the ability to generate high-quality EEG signals even in the face of such complexity, highlighting its capability to handle intricate EEG signal generation.

\subsubsection{Results and comparisons of data augmentation methods}
We compare the enhancement effect of our proposed method with the direct generated data incorporation method, and the results are shown in Table \ref{Table2}.

\begin{figure}[!ht]
    \centering
    \subfigure[The test accuracy of EEGNet]{
        \includegraphics[width=0.220\textwidth]{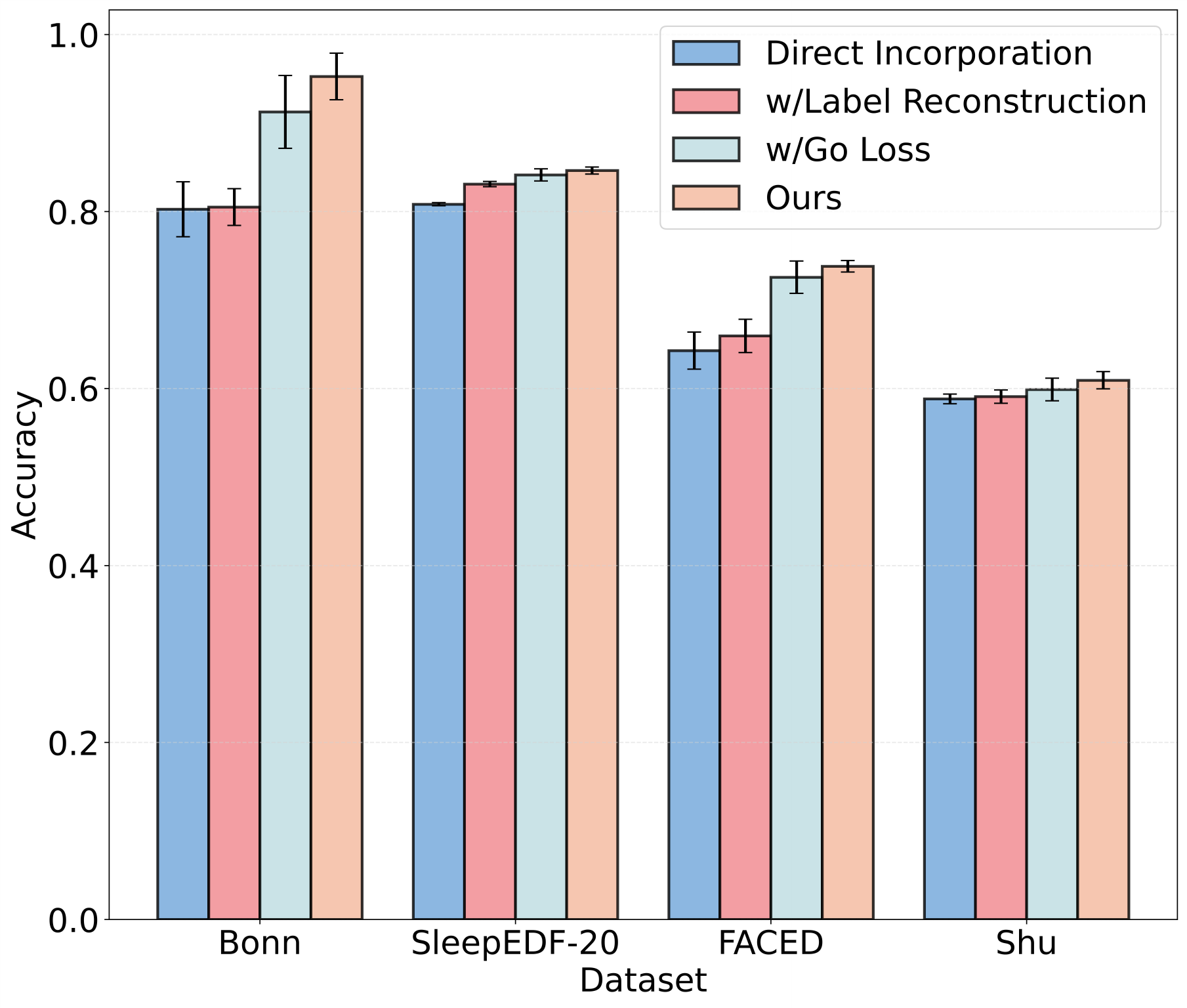}
    }
    \hfill
    \subfigure[The test accuracy of EEG-Conformer]{
        \includegraphics[width=0.220\textwidth]{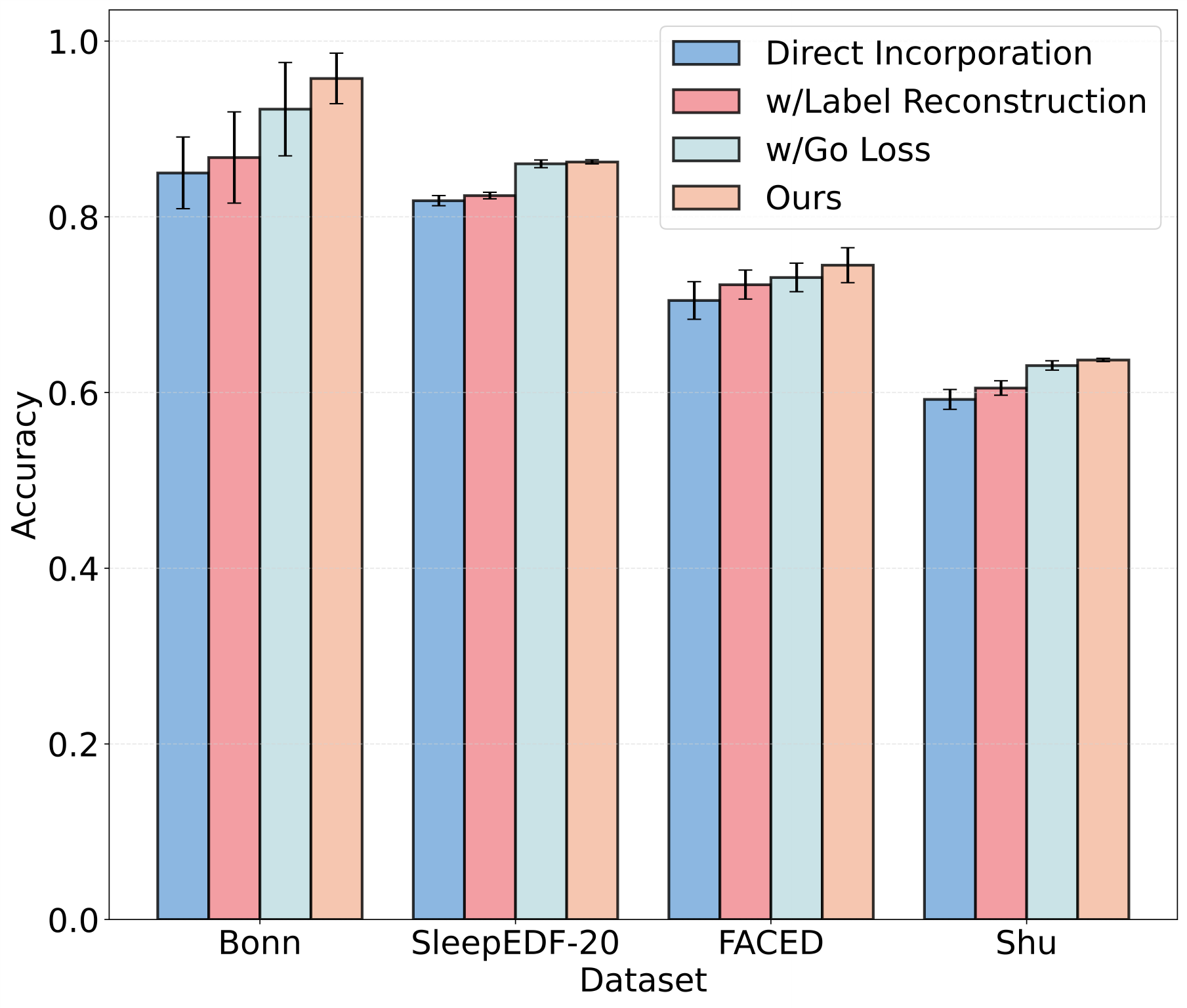}
    }
    \caption{Ablation on Label Reconstruction and Go Loss}
    \label{Figure4}
\end{figure}
It is obvious from Table \ref{Table2} that our proposed method can effectively enhance the EEG classification network, and the enhancement effect reaches the best on all the datasets. 

From Table \ref{Table2}, it is evident that our proposed enhancement methods, EEGNet and EEG-Conformer, yielded substantial performance improvements across four distinct datasets. Specifically, the maximum enhancements were observed as follows: 14.00\% on the Bonn dataset, 6.38\% on the SleepEDF-20 dataset, 9.42\% on the FACED dataset, and 2.5\% on the Shu dataset.

The experimental results indicate that our method is not reliant on specific network architectures or task types. It is capable of consistently improving the performance of classification networks without altering any existing network structures, demonstrating the generality and effectiveness of our approach.

\subsubsection{Ablation study on the proposed Data augmentation method}
To assess the impact of Label Reconstruction and Go Loss $\mathcal{L}_{GO}$ , we performed experiments by removing each module from our proposed data augmentation method to analyze their individual effects on EEG classification network performance. As demonstrated in Figure \ref{Figure4}, both Label Reconstruction and Go Loss $\mathcal{L}_{GO}$ individually improved the network's performance. Additionally, their combined use resulted in a greater enhancement compared to either module used alone.

\subsubsection{Ablation study on MSC and DFSI Modules}
We conducted an ablation study on the proposed Multi-Scale Convolution (MSC) module and Dynamic Fourier Spectrum Information (DFSI) module, calculating the FID for the data generated by the models after individually removing these two modules. The results are presented in Table \ref{Table3}.

\begin{table}[!ht]
\caption{Ablation Study on MSC and DFSI Modules with FID Evaluation.}
\centering
\scalebox{1}{%
\begin{tabular}{@{}cccc@{}}
\toprule
Dataset     & Ours          & w/o MSC & w/o DFSI      \\ \midrule
Bonn        & \textbf{3.81} & 5.58    & 4.86          \\
SleepEDF-20 & \textbf{5.25} & 6.46    & 12.26         \\
FACED       & \textbf{0.55} & 1.73    & 1.76          \\
Shu         & 0.24          & 0.23    & \textbf{0.20} \\ \bottomrule
\end{tabular}%
}
\label{Table3}
\end{table}

From Table \ref{Table3}, it can be observed that the proposed MSC and DFSI modules have shown improvement in the quality of generated data on most datasets. Concurrently, it has been observed that on the Shu dataset, there is no improvement in the quality of generated data with the proposed modules. This may be attributed to the fact that the Shu dataset has a relatively simplistic spectral composition, whereas the MSC and DFSI modules were designed to address the spectral diversity characteristics of EEG signals. Consequently, in this scenario, they may not exert a significant enhancing effect.

\section{Conclusion}
In this paper we propose a Transformer-based denoising diffusion probabilistic model and a generated data-based data augmentation method. The diffusion model is capable of generating high-quality EEG signals, surpassing existing EEG signal generative models in quality. The data augmentation method has been extensively validated on EEG datasets from multiple different tasks, and experimental results demonstrate the universality and effectiveness of our proposed methods. Our model is currently validated only on the EEG task and has not yet been tested on other time series tasks, and we will continue to investigate its effectiveness on a wider range of time series tasks in the future.

\bibliography{ref}
\newpage
\clearpage
\section{Appendix}

\subsection{Datasets Introduction}

\textbf{Bonn Dataset} The Bonn dataset is composed of EEG data from 5 healthy individuals and 5 patients with epilepsy, totaling 5 subsets, which are F, S, N, Z, and O. The Bonn dataset is a single-channel dataset, where each subset contains 100 data segments. The duration of each data segment is 23.6 seconds with a sampling frequency of 173.61Hz, a resolution of 12 bits, and a total of 4,097 signal points. During the data cutting process, noise signals such as myogenic and ocular artifacts have been removed. The data Z and O are scalp EEGs, collected from 5 healthy individuals, forming the control group. The segments in Z are EEGs when the subjects have their eyes open, and the segments in O are EEGs when the subjects have their eyes closed. The data N, F, and S are intracranial EEGs, collected from 5 patients who have been diagnosed preoperatively. N and F are collected during the interictal phase of epilepsy, and S is collected during the ictal phase. To facilitate model processing, we have truncated the length of this dataset to 4,096 signal points.

\textbf{FACED Dataset} The research group led by Zhang Dan from the Department of Psychology at Tsinghua University has released an EEG dataset for affective computing with fine-grained emotion categories, known as the Finer-grained Affective Computing EEG Dataset (FACED Dataset). The dataset encompasses 28 video segments of 32-channel EEG activity from 123 subjects as they watched, covering nine different emotions. These include four positive emotions (amusement, inspiration, happiness, and tenderness) defined according to the latest advancements in positive psychology research, four classical negative emotions (anger, disgust, fear, and sadness), as well as a neutral emotion. The FACED dataset provides a more refined description of emotions by subdividing the fine categories of positive and negative emotions, coupled with a large sample of subjects. This offers researchers the opportunity to better extract common emotional EEG characteristics across individuals, further advancing the research in the application of EEG-based affective computing. To expedite the training of the generative model and reduce GPU memory usage, we selected the Fp1, Fp2, and Fz electrodes from the dataset based on preliminary experiments.

\textbf{Shu Dataset} The motor imagery dataset experiment is divided into three phases. In the first phase, from 0 to 2 seconds, is the resting preparation period, during which subjects can rest, engage in minor physical activities, and blink. The second phase, from 2 to 4 seconds, is the cue phase. In this stage, an animation of the movement of the left or right hand will appear on the monitor, signaling to the subject the hand movement task they are about to perform. The third phase, from 4 to 8 seconds, is the MI (Motor Imagery) process. During this period, subjects perform the MI task of hand movement as prompted by the arrows on the monitor. In this phase, the EEG device collects and stores the signals. After completing one session, 100 trials are obtained. In the experiment, five separate sessions were conducted for each subject, with each session taking place every 2 to 3 days, so that each subject would ultimately have 500 trials across the 5 sessions. To expedite the training of the generative model and reduce GPU memory usage, we selected the Fp1, Fp2, and Fz electrodes from the dataset based on preliminary experiments.

\textbf{SleepEDF-20 Dataset} The SleepEDF-20 dataset was obtained from PhysioBank. It comprises data files from 20 subjects who participated in two studies. The first study involved polysomnographic recordings (SC files) to investigate the effects of age on sleep, focusing on healthy participants aged between 25 and 101 years. The second study involved sleep telemetry (ST files) to examine the effects of temazepam on the sleep of 22 Caucasian males and females while they were using other medications.For both datasets, each PSG (polysomnography) file contains two EEG channels (Fpz-Cz, Pz-Oz) sampled at 100 Hz, one EOG channel, and one chin EMG channel. Based on previous research, we utilized data from the sleep box study and employed a single Fpz-Cz channel as the input for various models in our experiments.

\subsection{Generated EEG Signals and Original EEG Signals Display}
Figures 1 to 3 depict the comparison between the EEG signals generated by our model and the original EEG signals, where Figures 1 present the images of single-channel EEG signals, while Figures 2 and 3 showcase the images of multi-channel EEG signals. It is evident from Figures 1 to 3 that our model is capable of generating EEG signals that closely resemble the original data, regardless of whether the signals are from a single channel or multiple channels. This demonstrates that our model possesses strong signal generation capabilities and exhibits commendable generalization performance.

\begin{figure}[!h]
    \centering
    \includegraphics[width=0.45\textwidth]{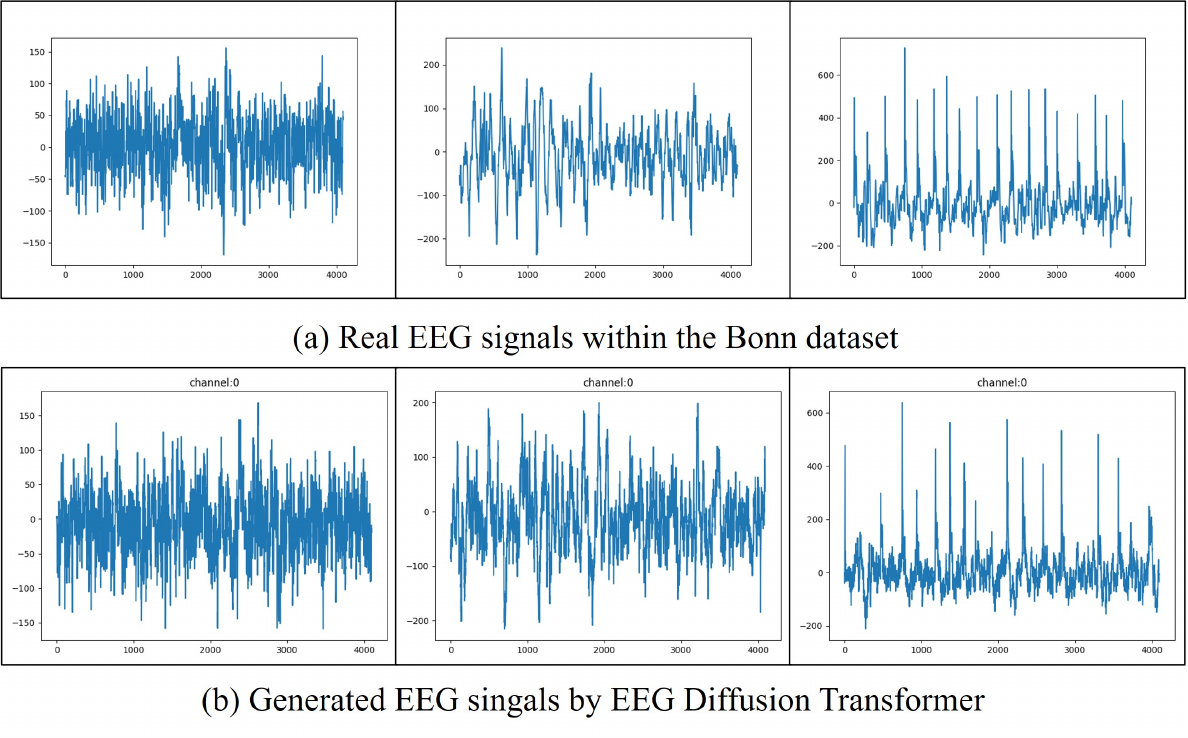}
    \caption{Real EEG signals and generated EEG signals on Bonn dataset.}
    \label{fig:Bonn}
\end{figure}
\newpage

\begin{figure}[h]
    \centering
    \includegraphics[width=0.45\textwidth]{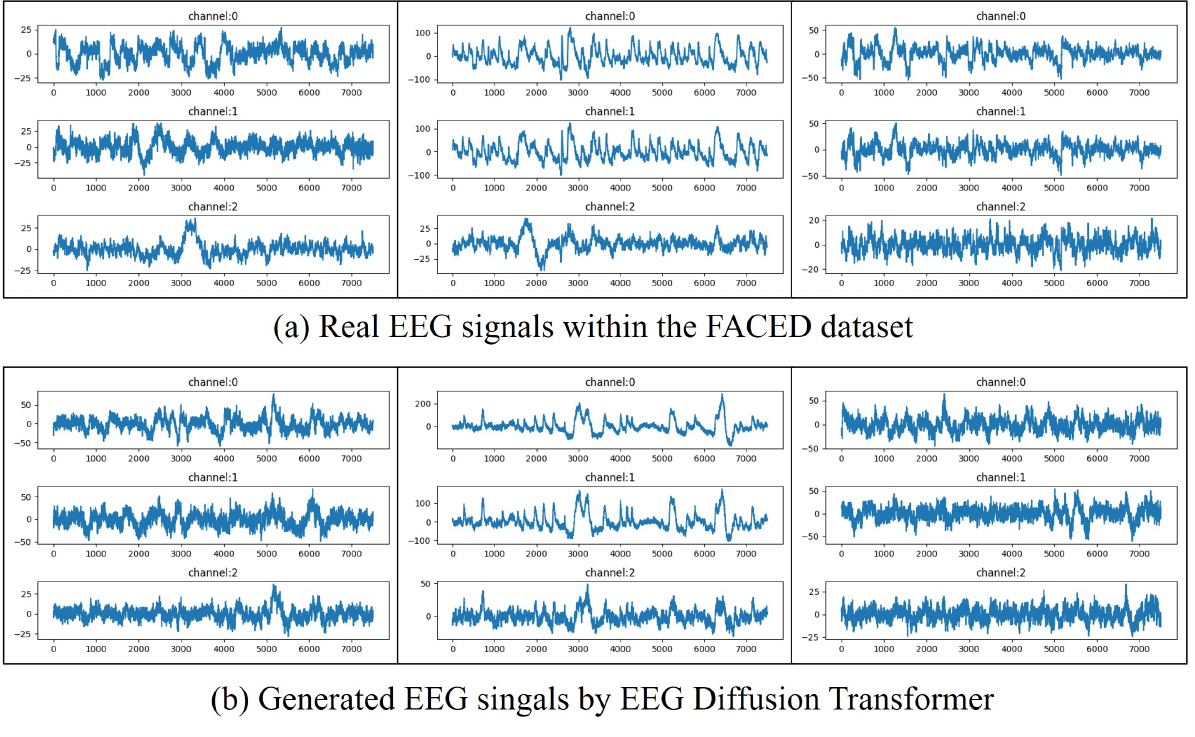}
    \caption{Real EEG signals and generated EEG signals on FACED dataset}
    \label{fig:FACED}

\end{figure}

\begin{figure}[h]
    \centering
    \includegraphics[width=0.45\textwidth]{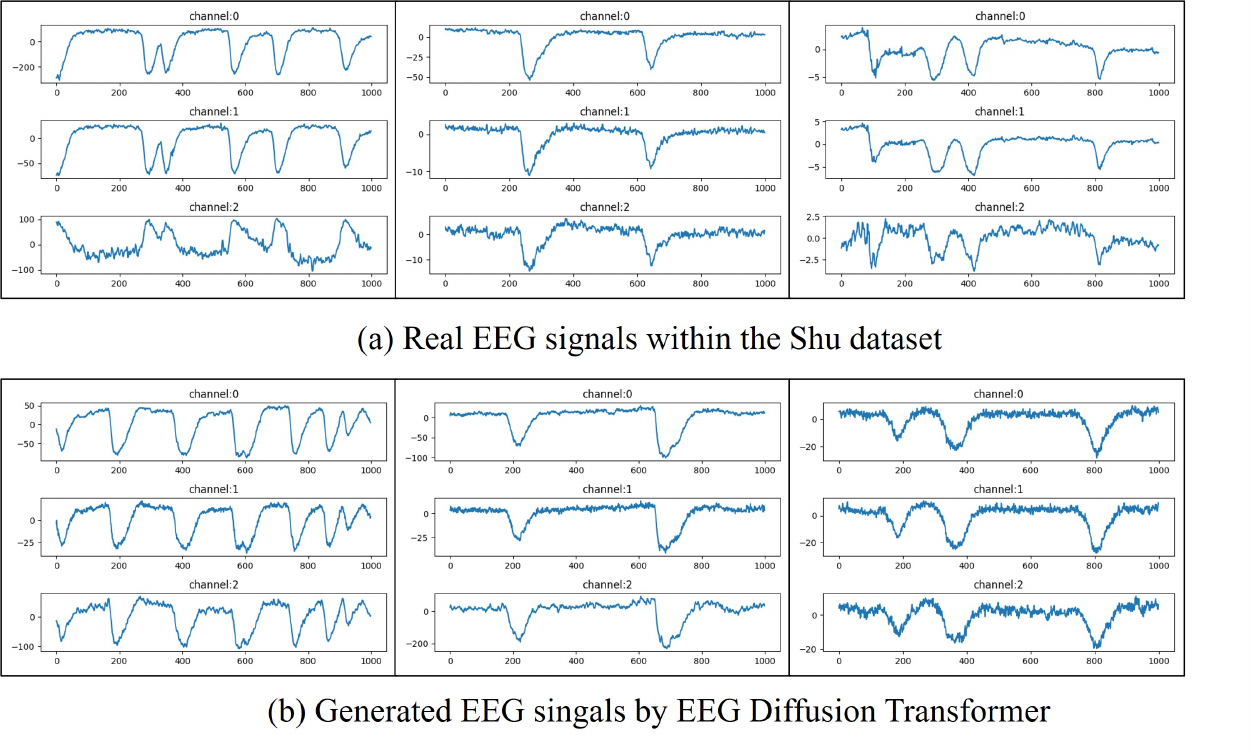}
    \caption{Real EEG signals and generated EEG signals on Shu dataset.}
    \label{fig:Shu}
\end{figure}

\end{document}